\documentclass[amsmath,amssymb,superscriptaddress,twocolumn,pra]{revtex4-1}
\usepackage{braket,hyperref,graphicx}
\pdfoutput=1
\newcommand{\abs}[1]{\left\lvert{#1}\right\rvert}
\DeclareMathOperator{\tr}{Tr}

\hypersetup{pdfauthor={Matthew F. Pusey, Jonathan Barrett and Terry Rudolph},pdftitle={On the reality of the quantum state}}
\begin{document}
\title{On the reality of the quantum state}

\author{Matthew F. Pusey}
\email{m@physics.org}
\affiliation{Department of Physics, Imperial College London, Prince Consort Road, London SW7 2AZ, United Kingdom}

\author{Jonathan Barrett}
\affiliation{Department of Mathematics, Royal Holloway, University of London, Egham Hill, Egham
TW20 0EX, United Kingdom}

\author{Terry Rudolph}
\affiliation{Department of Physics, Imperial College London, Prince Consort Road, London SW7 2AZ, United Kingdom}

\date{April 11, 2012}

\begin{abstract}
Quantum states are the key mathematical objects in quantum theory. It is therefore surprising that physicists have been unable to agree on what a quantum state truly represents. One possibility is that a pure quantum state corresponds directly to reality. However, there is a long history of suggestions that a quantum state (even a pure state) represents only knowledge or information about some aspect of reality. Here we show that any model in which a quantum state represents mere information about an underlying physical state of the system, and in which systems that are prepared independently have independent physical states, must make predictions which contradict those of quantum theory.
\end{abstract}
\maketitle

At the heart of much debate concerning quantum theory lies the quantum state. Does the wave function correspond directly to some kind of physical wave? If so, it is an odd kind of wave, since it is defined on an abstract configuration space, rather than the three-dimensional space we live in. Nonetheless, quantum interference, as exhibited in the famous two-slit experiment, appears most readily understood by the idea that it is a real wave that is interfering.  Many physicists and chemists concerned with pragmatic applications of quantum theory successfully treat the quantum state in this way.

Many others have suggested that the quantum state is something less than real \cite{epr,popper,ballen,peierls,jaynes,zeilinger,quantbays,toy}. In particular, it is often argued that the quantum state does not correspond directly to reality, but represents an experimenter's knowledge or information about some aspect of reality. This view is motivated by, amongst other things, the collapse of the quantum state on measurement. If the quantum state is a real physical state, then collapse is a mysterious physical process, whose precise time of occurrence is not well-defined. From the `state of knowledge' view, the argument goes, collapse need be no more mysterious than the instantaneous Bayesian updating of a probability distribution upon obtaining new information.

The importance of these questions was eloquently stated by Jaynes:
\begin{quotation}
\emph{But our present [quantum mechanical] formalism is not purely epistemological; it is a peculiar mixture describing in part realities of Nature, in part incomplete human information about Nature --- all scrambled up by Heisenberg and Bohr into an omelette that nobody has seen how to unscramble. Yet we think that the unscrambling is a prerequisite for any further advance in basic physical theory. For, if we cannot separate the subjective and objective aspects of the formalism, we cannot know what we are talking about; it is just that simple.\cite{jaynesquote2}}
\end{quotation}

Here we present a no-go theorem: if the quantum state merely represents information about the real physical state of a system, then experimental predictions are obtained which contradict those of quantum theory. The argument depends on few assumptions. One is that a system \emph{has} a ``real physical state'' -- not necessarily completely described by quantum theory, but objective and independent of the observer. This assumption only needs to hold for systems that are isolated, and not entangled with other systems. Nonetheless, this assumption, or some part of it, would be denied by instrumentalist approaches to quantum theory, wherein the quantum state is merely a calculational tool for making predictions concerning macroscopic measurement outcomes. The other main assumption is that systems that are prepared independently have independent physical states.

In order to make some of these notions more precise, let us begin by considering the classical mechanics of a point particle moving in one dimension. At a given moment of time, the physical state of the particle is completely specified by its position $x$ and momentum $p$, and hence corresponds to a point $(x,p)$ in a two-dimensional phase space. Other physical properties are either fixed, such as mass or charge, or are functions of the state, such as energy $H(x,p)$. Viewing the fixed properties as constant functions, let us \emph{define} ``physical property'' to mean some function of the physical state.

Sometimes, the exact physical state of the particle might be uncertain, but there is nonetheless a well-defined probability distribution $\mu(x,p)$. Although $\mu(x,p)$ evolves in a precise manner according to Liouville's equation, it does not directly represent reality. Rather, $\mu(x,p)$ is a \emph{state of knowledge}: it represents an experimenter's uncertainty about the physical state of the particle. 

Now consider a quantum system. The hypothesis is that the quantum state is a state of knowledge, representing uncertainty about the real physical state of the system. Hence assume some theory or model, perhaps undiscovered, which associates a physical state $\lambda$ to the system. If a measurement is performed, the probabilities for different outcomes are determined by $\lambda$. If a quantum system is prepared in a particular way, then quantum theory associates a quantum state (assume for simplicity that it is a pure state) $\ket{\psi}$. But the physical state $\lambda$ need not be fixed uniquely by the preparation -- rather, the preparation results in a physical state $\lambda$ according to some probability distribution $\mu_\psi(\lambda)$.

\begin{figure}
  \includegraphics[width=\columnwidth]{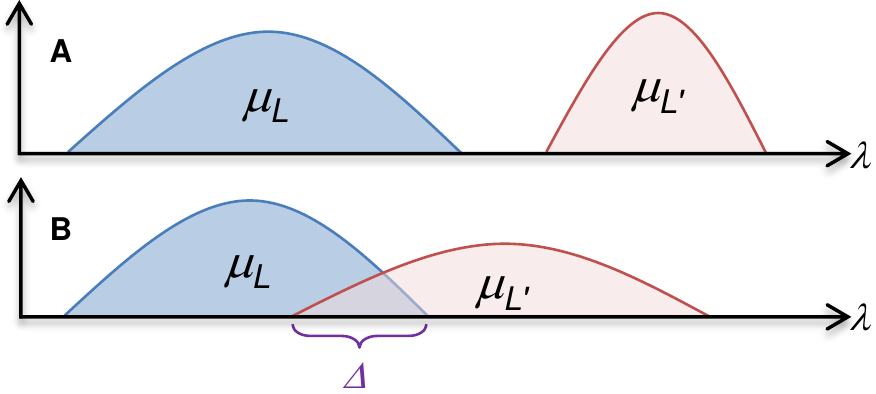}
  \caption{Our definition of a physical property is illustrated. Consider a collection, labelled by $L$, of probability distributions $\{\mu_L(\lambda)\}$. $\lambda$ denotes a system's physical state. If every pair of distributions are disjoint, as in {\bf a}, then the label $L$ is uniquely fixed by $\lambda$ and we call it a physical property. If, however, $L$ is not a physical property, then there exists a pair of labels $L, L'$ with distributions that both assign positive probability to some overlap region $\Delta$, as in {\bf b}. A $\lambda$ from $\Delta$ is consistent with either label.}

  \label{fig1}
\end{figure}

Given such a model, Harrigan and Spekkens\cite{nic} give a precise meaning to the idea that a quantum state corresponds directly to reality or represents only information. To explain this, the example of the classical particle is again useful. Here, if an experimenter knows only that the system has energy $E$, and is otherwise completely uncertain, the experimenter's knowledge corresponds to a distribution $\mu_E(x,p)$ uniform over all points in phase space with $H(x,p)=E$. Seeing as the energy is a physical property of the system, different values of the energy $E$ and $E'$ correspond to disjoint regions of phase space, hence the distributions $\mu_E(x,p)$ and $\mu_{E'}(x,p)$ have disjoint supports. On the other hand, if two probability distributions $\mu_L(x,p)$ and $\mu_{L'}(x,p)$ have overlapping supports, i.e. there is some region $\Delta$ of phase space where both distributions are non-zero, then the labels $L$ and $L'$ cannot refer to a physical property of the system. See Figure~\ref{fig1}.

Similar considerations apply in the quantum case. Suppose that, for any pair of distinct quantum states $\ket{\psi_0}$ and $\ket{\psi_1}$, the distributions $\mu_0(\lambda)$ and $\mu_1(\lambda)$ do not overlap: then, the quantum state $|\psi\rangle$ can be inferred uniquely from the physical state of the system and hence satisfies the above definition of a physical property. Informally, every detail of the quantum state is ``written into'' the real physical state of affairs. But if $\mu_0(\lambda)$ and $\mu_1(\lambda)$ overlap for at least one pair of quantum states, then $|\psi\rangle$ can justifiably be regarded as ``mere'' information.

Our main result is that for distinct quantum states $\ket{\psi_0}$ and $\ket{\psi_1}$, if the distributions $\mu_0(\lambda)$ and $\mu_1(\lambda)$ overlap (more precisely: if $\Delta$, the intersection of their supports, has non-zero measure) then there is a contradiction with the predictions of quantum theory. We present first a simple version of the argument, which works when $|\braket{\psi_0|\psi_1}| = 1/\sqrt{2}$. Then the argument is extended to arbitrary $\ket{\psi_0}$ and $\ket{\psi_1}$. Finally, we present a more formal version of the argument which works even in the presence of experimental error and noise.

Consider two methods of preparing a quantum system, corresponding to quantum states $\ket{\psi_0}$ and $\ket{\psi_1}$, with $|\braket{\psi_0|\psi_1}| = 1/\sqrt{2}$. Choose a basis of the Hilbert space so that $\ket{\psi_0} = \ket{0}$ and $\ket{\psi_1} = \ket{+} = (\ket{0} + \ket{1})/\sqrt 2$. In order to derive a contradiction, suppose that the distributions $\mu_0(\lambda)$ and $\mu_1(\lambda)$ overlap. Then there exists $q>0$ such that preparation of either quantum state results in a $\lambda$ from the overlap region $\Delta$ with probability at least $q$.

\begin{figure}
  \includegraphics[width=\columnwidth]{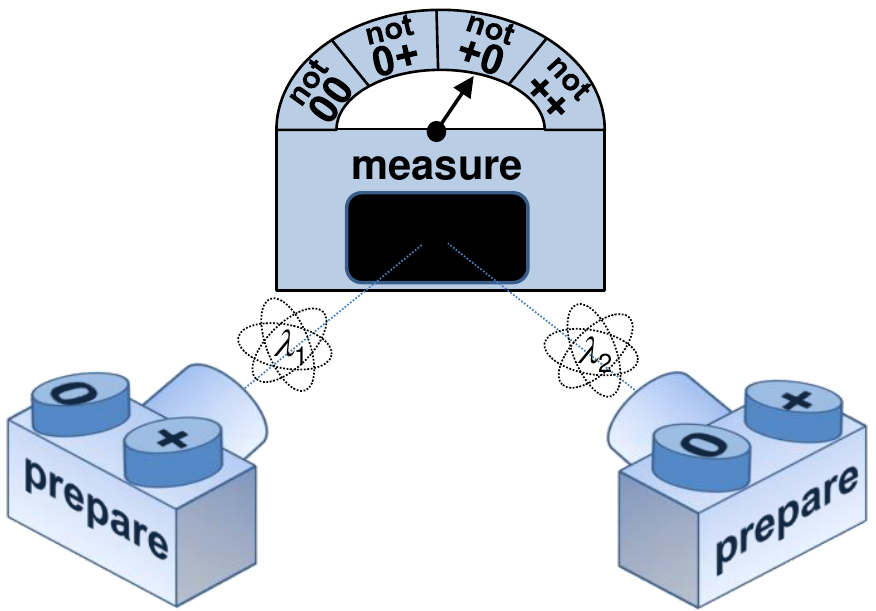}
  \caption{Two systems are prepared independently. The quantum state of each, determined by the preparation method, is either $\ket{0}$ or $\ket{+}$. The two systems are brought together and measured. The outcome of the measurement can only depend on the physical states of the two systems at the time of measurement.}
  \label{fig2}
\end{figure}

Now consider two systems whose physical states are uncorrelated. This can be achieved, for example, by constructing and operating two copies of a preparation device independently. Each system can be prepared such that its quantum state is either $|\psi_0\rangle$ or $|\psi_1\rangle$, as illustrated in Figure~\ref{fig2}. With probability $q^2>0$ it happens that the physical states $\lambda_1$ and $\lambda_2$ are both from the overlap region $\Delta$.
This means that the physical state of the two systems is compatible with any of the four possible quantum states $\ket{0}\otimes\ket{0}$, $\ket{0}\otimes\ket{+}$, $\ket{+}\otimes\ket{0}$ and $\ket{+}\otimes\ket{+}$.

The two systems are brought together and measured. The measurement is an entangled measurement, which projects onto the four orthogonal states:
\begin{eqnarray}\label{twoboxbasis}
\ket{\xi_1}&=&\tfrac{1}{\sqrt{2}}(\ket{0}\otimes\ket{1}+\ket{1}\otimes\ket{0}), \nonumber\\
\ket{\xi_2}&=&\tfrac{1}{\sqrt{2}}(\ket{0}\otimes\ket{-}+\ket{1}\otimes\ket{+}), \nonumber\\
\ket{\xi_3}&=&\tfrac{1}{\sqrt{2}}(\ket{+}\otimes\ket{1}+\ket{-}\otimes\ket{0}), \nonumber\\
\ket{\xi_4}&=&\tfrac{1}{\sqrt{2}}(\ket{+}\otimes\ket{-}+\ket{-}\otimes\ket{+}),
\end{eqnarray}
where $\ket{-}=(\ket{0}-\ket{1})/\sqrt{2}$.
The first outcome is orthogonal to $\ket{0}\otimes\ket{0}$, hence quantum theory predicts that this outcome has probability zero when the quantum state is $\ket{0}\otimes\ket{0}$. Similarly, outcome $\ket{\xi_2}$ has probability zero if the state is $\ket{0}\otimes\ket{+}$, $\ket{\xi_3}$ if $\ket{+}\otimes\ket{0}$, and $\ket{\xi_4}$ if $\ket{+}\otimes\ket{+}$.
This leads immediately to the desired contradiction. At least $q^2$ of the time, the measuring device is uncertain which of the four possible preparation methods was used, and on these occasions it runs the risk of giving an outcome that quantum theory predicts should occur with probability $0$. Importantly, we have needed to say nothing about the value of $q$ \emph{per se} to arrive at this contradiction.

We have shown that the distributions for $\ket{0}$ and $\ket{+}$ cannot overlap. If the same can be shown for any pair of quantum states $\ket{\psi_0}$ and $\ket{\psi_1}$, then the quantum state can be inferred uniquely from $\lambda$. In this case, the quantum state is a physical property of the system.

For any pair of distinct non-orthogonal states $\ket{\psi_0}$ and $\ket{\psi_1}$, a basis of the Hilbert space can be chosen such that
\begin{align}
\ket{\psi_0} &= \cos (\theta/2) \ket{0} + \sin (\theta/2) \ket{1}\nonumber\\
\ket{\psi_1} &= \cos(\theta/2) \ket{0} - \sin (\theta/2) \ket{1},
\end{align}
with $0 < \theta < \pi/2$. These states span a two-dimensional subspace of the Hilbert space. We can restrict attention to this subspace and from hereon, without loss of generality, treat the systems as qubits. As above, suppose that there is a probability at least $q>0$ that the physical state of the system after preparation is compatible with either preparation method having been used, that is, the resulting $\lambda$ is in $\Delta$.

A contradiction is obtained when $n$ uncorrelated systems are prepared, where $n$ will be fixed shortly. Depending on which of the two preparation methods is used each time, the $n$ systems are prepared in one of the quantum states:
\begin{equation}
\ket{\Psi ( x_1 \ldots x_n )} = \ket{\psi_{x_1}}\otimes\cdots\otimes\ket{\psi_{x_{n-1}}}\otimes\ket{\psi_{x_n}},
\end{equation}
where $x_i\in\{0,1\}$, for each $i$. Since the preparations are independent, there is a probability at least $q^n$ that the complete physical state of the systems emerging from the devices is compatible with any one of these $2^n$ quantum states. The contradiction is obtained if there is a joint measurement on the $n$ systems such that each outcome has probability zero on at least one of the $\ket{\Psi(x_1 \ldots x_n)}$. (This type of measurement was first introduced in a different context by Caves, Fuchs and Schack \cite{compat}; in their terminology, the existence of such a measurement shows the states are \emph{Post-Peierls-incompatible}.)

\begin{figure}
  \includegraphics[width=\columnwidth]{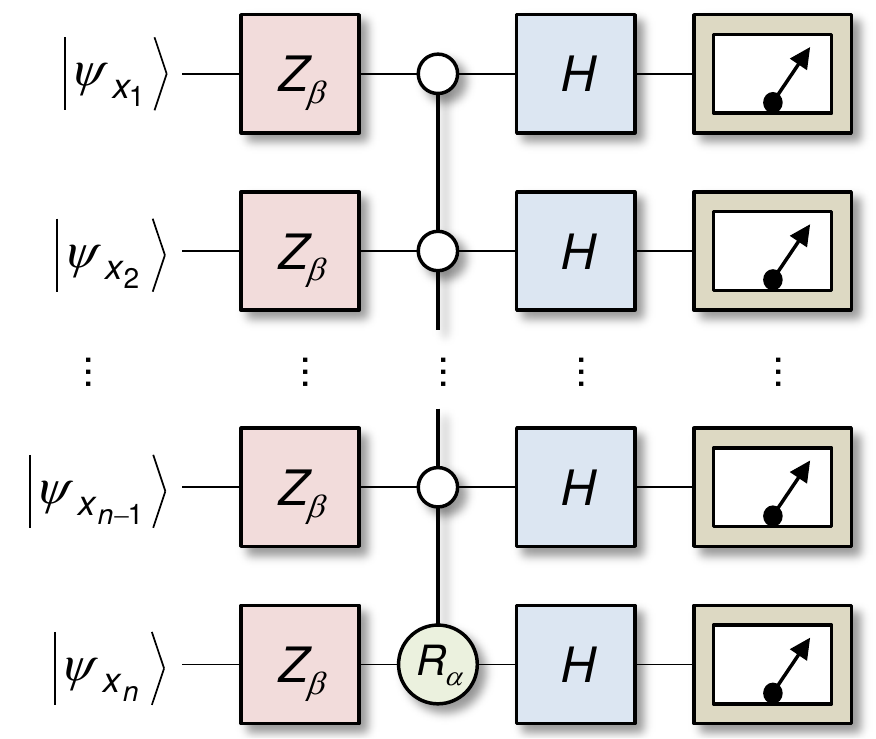}
  \caption{The main argument requires a joint measurement on $n$ qubits with the property that each outcome has probability zero on one of the input states. Such a measurement can be performed by implementing the quantum circuit shown, followed by a measurement of each qubit in the computational basis. The single qubit gates are given by
$Z_\beta =\ket{0}\bra{0}+e^{i\beta}\ket{1}\bra{1}$
and the Hadamard gate
$H =\ket{+}\bra{0}+\ket{-}\bra{1}$.
The entangling gate in the middle rotates the phase of only one state:
$R_\alpha\ket{00\ldots0}= e^{i\alpha}\ket{00\ldots0}$, leaving all other computational basis states unaffected.}
  \label{fig3}
\end{figure}

A suitable measurement is most easily described as a quantum circuit, followed by a measurement onto the $\{\ket{0},\ket{1}\}$ basis for each qubit. It is illustrated in Figure~\ref{fig3}.

The circuit is parameterized by two real numbers, $\alpha$ and $\beta$. In Appendix~\ref{eqnsol} it is shown that for any $0<\theta<\pi/2$, and for any $n$ chosen large enough that  $2^{1/n} - 1 \leq \tan (\theta/2)$, it is possible to choose $\alpha$ and $\beta$ such that the measurement has the desired feature: each outcome has, according to quantum theory, probability zero on one of the states $\ket{\Psi(x_1\ldots x_n)}$.

The presentation so far has been somewhat heuristic. We turn to a more formal statement of the result, including the possibility of experimental error. This is important because the argument so far uses the fact that quantum probabilities are sometimes exactly zero. It is important to have a version of the argument which is robust against small amounts of noise. Otherwise the conclusion -- that the quantum state is a physical property of a quantum system -- would be an artificial feature of the exact theory, but irrelevant to the real world and experimental test would be impossible.

Let us restate our assumptions more mathematically. First, assume a measure space $\Lambda$, understood as the set of possible physical states $\lambda$ that a system can be in. Preparation of the quantum state $\ket{\psi_i}$ is assumed to result a $\lambda$ sampled from a probability distribution $\mu_i(\lambda)$ over $\Lambda$. Second, assume that it is possible to prepare $n$ systems independently, with quantum states $\ket{\psi_{x_1}}, \dotsc, \ket{\psi_{x_n}}$, resulting in physical states $\lambda_1, \dotsc, \lambda_n$ distributed according to the product distribution
\begin{equation}
 \mu_{x_1}(\lambda_1)\mu_{x_2}(\lambda_2)\dotsm\mu_{x_n}(\lambda_n).
\end{equation}
Finally, assume that $\lambda_1, \dotsc, \lambda_n$ fixes the probability for the outcome $k$ of a measurement according to some probability distribution $p(k|\lambda_1, \dotsc, \lambda_n)$. The operational probabilities $p\left( k|\Psi(x_1\ldots x_n)\right)$  are given by
\begin{equation}
  \int_\Lambda \dotsm \int_\Lambda p(k|\lambda_1, \dotsc, \lambda_n)\mu_{x_1}(\lambda_1)\dotsm \mu_{x_n}(\lambda_n) d\lambda_1 \dotsm d\lambda_n.
\end{equation}

If an experiment is performed, it will be possible to establish with high confidence that the probability for each measurement outcome is within $\epsilon$ of the predicted quantum probability for some small $\epsilon > 0$. The final result relates $\epsilon$ to the \emph{total variation distance} \cite{probmet} between $\mu_0$ and $\mu_1$, defined by
\begin{equation}
D(\mu_0,\mu_1) = \frac12\int_{\Lambda} |\mu_0(\lambda) - \mu_1(\lambda)| \mathrm{d}\lambda.
\end{equation}
It is a measure of how easy it is to distinguish two probability distributions. If $D(\mu_0,\mu_1)=1$, then $\mu_0$ and $\mu_1$ are completely disjoint. In this case, the probability of $\lambda$ being compatible with both preparations ($q$ above) is zero. In Appendix~\ref{noiseversion} we show that if the probabilities predicted by a model are within $\epsilon$ of the quantum probabilities then
\begin{equation}
D(\mu_0,\mu_1) \geq 1 - 2\sqrt[n]{\epsilon},
\label{finalresult}
\end{equation}
for $2^{1/n} - 1 \leq \tan (\theta/2)$.
For small $\epsilon$, $D(\mu_0,\mu_1)$ is close to $1$. Hence a successful experiment would show that $\lambda$ is normally closely associated with only one of the two quantum states.

Performing an experiment to implement the circuit in Figure~\ref{fig3} for small values of $n$ is challenging but not unrealistic given current technology. While all the gates required have already been demonstrated at some point, our result requires such gates acting with high fidelity in a non post-selected fashion (this latter because otherwise the measuring device can use the extra freedom in the postselection to escape the zero-probability outcomes those times it is unsure of the preparation procedure).

In conclusion, we have presented a no-go theorem, which -- modulo assumptions -- shows that models in which the quantum state is interpreted as mere information about an objective physical state of a system cannot reproduce the predictions of quantum theory. The result is in the same spirit as Bell's theorem\cite{bell}, which states that no \emph{local} theory can reproduce the predictions of quantum theory. Both theorems need to assume that a system has a objective physical state $\lambda$ such that probabilities for measurement outcomes depend only on $\lambda$. But our theorem only assumes this for systems prepared in isolation from the rest of the universe in a quantum pure state. This is unlike Bell's theorem, which needs to assume the same thing for entangled systems. Neither theorem assumes underlying determinism.

Bell's theorem assumes that it is possible to make independent choices of measurement, and since local models which drop measurement independence can be constructed\cite{indep,indep2}, this assumption is necessary.  Somewhat analogously, models where the quantum state is not a physical property can be constructed by dropping our assumption of preparation independence\cite{ljbr}. Since both assumptions are very reasonable, it is not surprising that in both cases the models obtained by dropping them appear extremely contrived.

An important step towards the derivation of our result is the idea that the quantum state is physical if distinct quantum states correspond to non-overlapping distributions for $\lambda$. The precise formalisation of this idea appeared in Spekkens\cite{context} and in Harrigan and Spekkens\cite{nic}, and is also due to Hardy\cite{hpriv}. In the terminology of Harrigan and Spekkens, we have shown that \emph{$\psi$-epistemic} models cannot reproduce the predictions of quantum theory. The general notion that two distinct quantum states may describe the same state of reality, however, has a long history. For example, in a letter to Schroedinger containing a variant of the famous EPR (Einstein-Podolsky-Rosen) argument\cite{epr}, Einstein argues from locality to the conclusion that
\begin{quotation}
...for the same [real] state of [the system at] $B$ there are two (in general arbitrarily many) equally justified $\Psi_B$, which contradicts the hypothesis of a one-to-one or complete description of the real states. \cite{einquote}
\end{quotation}
In this version of the argument, Einstein really is concerned with the possibility that there are two distinct quantum states for the same reality. He is not concluding that there are two different states of reality corresponding to the same quantum state (which would be the more commonly understood notion of incompleteness associated with Einstein)."

Finally, what are the consequences if we simply accept both the assumptions and the conclusion of the theorem? If the quantum state is a physical property of a system then quantum collapse must correspond to a -- problematic and poorly defined -- physical process. If there is no collapse, on the other hand, then after a measurement takes place, the joint quantum state of the system and measuring apparatus is entangled and contains a component corresponding to each possible macroscopic measurement outcome. This would be unproblematic if the quantum state merely reflected a lack of information about which outcome occurred. But if the quantum state is a physical property of the system and apparatus, it is hard to avoid the conclusion that each macroscopically different component has a direct counterpart in reality.

On a related, but more abstract note, the quantum state has the striking property that the number of real parameters needed to specify it is exponential in the number of systems $n$. This is to be expected if the quantum state represents information but is -- to us -- very surprising if it has a direct image in reality. Note that in previous work, Hardy has shown that the set $\Lambda$ of physical states must have infinite cardinality \cite{baggage}, and Montina has shown that, given some assumptions about the underlying dynamics, the physical state must have at least as many real parameters as the quantum state \cite{montina2,montina}. Similar conclusions can be drawn from ideas in communication complexity \cite{complex}.  

For these reasons and others, many will continue to view the quantum state as representing information. One approach is to take this to be information about possible measurement outcomes, and \emph{not} about the objective state of a system \cite{QUBISM}. Another is to construct concrete models of reality wherein one or more of our assumptions fail.

\begin{acknowledgments}
We thank Koenraad Audenaert for code, and Lucien Hardy, Matt Leifer, and Rob Spekkens for useful discussions. All authors are supported by the EPSRC.
\end{acknowledgments}

\bibliography{nogo}

\appendix
\section{The measurement circuit}\label{eqnsol}

Consider a preparation device which can produce a quantum system in either the state $\ket{\psi_0}$, or the state $\ket{\psi_1}$. Suppose that $n$ copies of this device are used independently. Then there are $2^n$ possible joint states of the $n$ systems, depending on whether $\ket{\psi_0}$ or $\ket{\psi_1}$ was prepared each time. This section shows that for any distinct $\ket{\psi_0}$ and $\ket{\psi_1}$, if the number of systems $n$ is large enough, then there is a joint measurement of the $n$ systems with the following property:  each outcome has zero probability given one of the $2^n$ possible preparations.

Choose a basis  $\{\ket{0},\ket{1}\}$ such that 
\begin{equation}
\ket{\psi_0} = \cos\frac\theta2\ket0 + \sin\frac\theta2\ket1,
\end{equation}
\begin{equation}
  \ket{\psi_1} = \cos\frac\theta2\ket0 - \sin\frac\theta2\ket1,
\end{equation}
where $\abs{\Braket{\psi_0|\psi_1}}^2 = \cos^2(\theta)$. By restricting attention to the subspace spanned by $\ket{\psi_0}$ and $\ket{\psi_1}$, we can without loss of generality take the quantum systems to be qubits. For reasons seen below, choose $n$ large enough that
\begin{equation}
2\arctan\left(2^{1/n} - 1\right) \leq \theta.  \label{ppreq}
\end{equation}


The circuit consists of a unitary rotation $Z_{\beta}$ applied to each qubit, followed by an entangling gate $R_{\alpha}$, followed by a Hadamard gate applied to each qubit. The initial rotation is given by
\begin{equation}
Z_\beta = \begin{pmatrix}1 & 0 \\ 0 & e^{i\beta}\end{pmatrix}.
\end{equation}
The $n$-qubit gate $R_\alpha$ is defined via its action on the computational basis states. Let $R_\alpha\ket{0\dotsm0} = e^{i\alpha}\ket{0\dotsm0}$, and let $R_\alpha$ act as the identity on all other computational basis states. Finally, the Hadamard gate corresponds to the unitary operation
\begin{equation}
H = \frac{1}{\sqrt 2}\begin{pmatrix}1 & 1 \\ 1 & -1\end{pmatrix}.
\end{equation}
The action of the circuit is given by $U_{\alpha,\beta} = H^{\otimes n} R_\alpha {Z_\beta}^{\otimes n}$.
The measurement procedure consists of the unitary evolution $U_{\alpha, \beta}$ (for a particular choice of $\alpha$ and $\beta$ discussed below), followed by a measurement of each qubit in the $\{\ket{0},\ket{1}\}$ basis.

Let $x_i$ be 0 (1) if the $i$th system is prepared in the state $\ket{\psi_0}$ ($\ket{\psi_1}$), and write $\vec{x} = (x_1,\ldots, x_n)$. Before the circuit is applied, the joint state of the $n$ systems is a direct product 
\begin{equation}\label{jointstate}
\ket{\Psi(\vec{x})} = \ket{\psi_{x_1}}\otimes\cdots\otimes\ket{\psi_{x_n}}. 
\end{equation}
If the initial preparation is $\ket{\Psi(x_1,\ldots, x_n)}$, then the probability of the measurement outcome corresponding to the basis state $\ket{x_1 \ldots x_n}$ is the squared absolute value of
\begin{widetext}
\begin{align}
& \bra{x_1 \ldots x_n} H^{\otimes n}R_\alpha Z_\beta^{\otimes n}\ket{\psi_{x_1}}\otimes \cdots\otimes \ket{\psi_{x_n}}\nonumber \\
& = \frac{1}{\sqrt{2^n}}\left(\sum_{\vec{z}} (-1)^{\vec{x} . \vec{z}} \bra{\vec{z}}\right) R_\alpha Z_\beta^{\otimes n} \ket{\psi_{x_1}}\otimes \cdots\otimes \ket{\psi_{x_n}} \nonumber \\
& = \frac{1}{\sqrt{2^n}}\left(e^{i\alpha}\bra{0\dotsm0} + \sum_{\vec{z}\ne 00\cdots 0} (-1)^{\vec{x} . \vec{z}} \bra{\vec{z}}\right) Z_\beta^{\otimes n}\ket{\psi_{x_1}}\otimes \cdots\otimes \ket{\psi_{x_n}} \nonumber \\
& = \frac{1}{\sqrt{2^n}}\left(e^{i\alpha}\bra{0\dotsm0} + \sum_{\vec{z}\ne 00\cdots 0} (-1)^{\vec{x} . \vec{z}} \bra{\vec{z}} \right) \bigotimes_{i=1}^n \left(\cos\frac\theta2\ket0 + (-1)^{x_i} e^{i\beta}\sin\frac\theta2 \ket{1}\right)\nonumber\\
& = \frac{1}{\sqrt{2^n}}\left(\left(\cos\frac\theta2\right)^n e^{i\alpha} + \sum_{\vec{z}\ne 00\cdots 0} (-1)^{\vec{x} . \vec{z}} \left(\cos\frac\theta2\right)^{n-|\vec{z}|}\left(\sin\frac\theta2\right)^{|\vec{z}|} e^{i |\vec{z}| \beta} (-1)^{\vec{x}.\vec{z}}\right) \nonumber \\
& = \frac{1}{\sqrt{2^n}}\left(\left(\cos\frac\theta2\right)^n e^{i\alpha} + \sum_{k=1}^n \left(\begin{array}{c} n \\ k \end{array}\right) \left(\cos\frac\theta2\right)^{n-k}\left(\sin\frac\theta2\right)^{k} e^{i k \beta}\right) \nonumber \\
& = \frac{1}{\sqrt{2^n}}\left(\cos\frac\theta2\right)^n \left(e^{i\alpha} + \left(1+e^{i\beta}\tan\frac\theta2\right)^n - 1\right).
\end{align}
\end{widetext}
In the fifth line, $|\vec{z}| = \sum_i z_i$.

Finally, we show that for any $\theta$ with $2\arctan\left(2^\frac{1}{n} - 1\right) \leq \theta \leq \frac\pi2$, the angles $\alpha$ and $\beta$ can be chosen so that
\begin{equation}
e^{i\alpha} + \left(1+e^{i\beta}\tan\frac\theta2\right)^n - 1 = 0,
\end{equation}
and hence the probability is zero as required.
Rearranging, the required $\alpha$ will always exist (and be easy to find) provided there exists a $\beta$ with
\begin{equation}
  \abs{1 - \left(1+e^{i\beta}\tan\frac\theta2\right)^n} = 1.\label{betaeq}
\end{equation}
\begin{figure}
  \begin{center}
\includegraphics[scale=0.7]{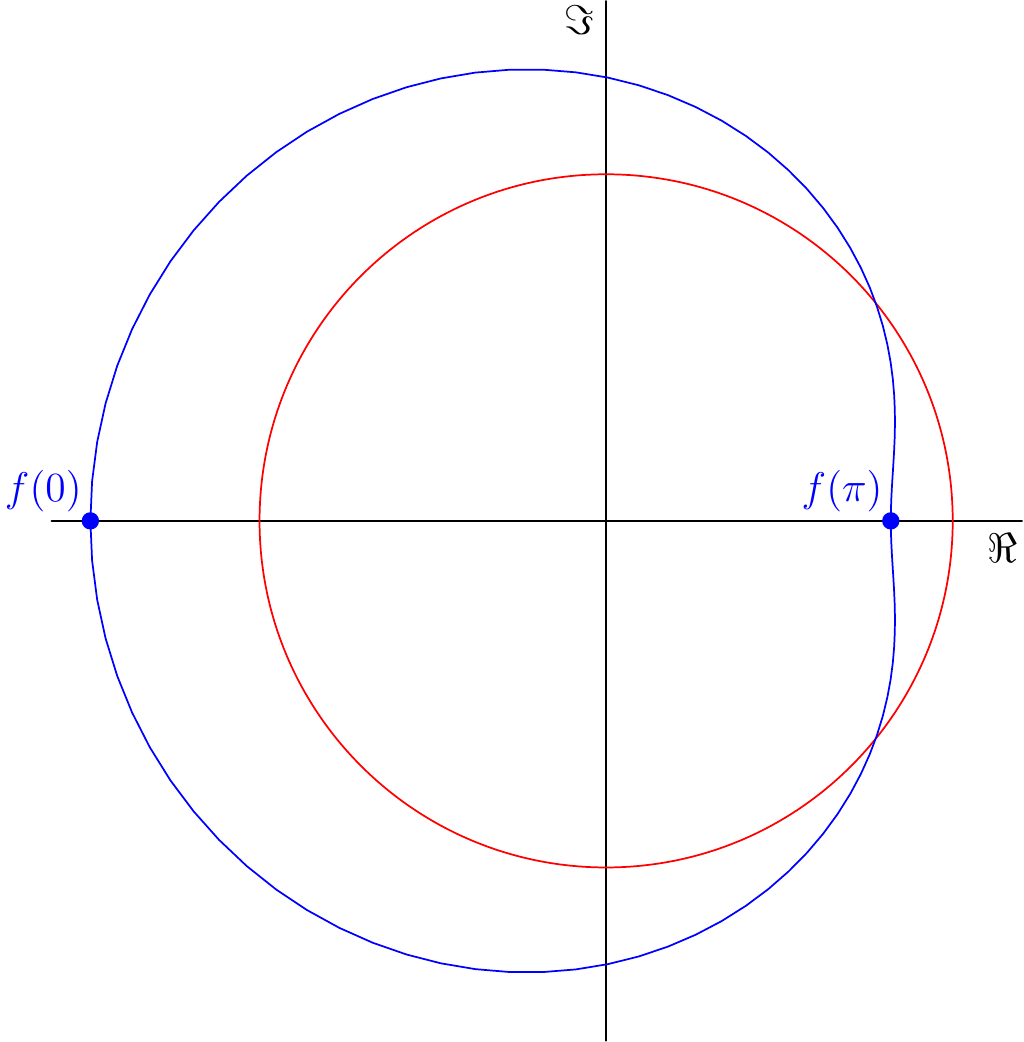}
  \end{center}
  \caption{Graph of $f(\beta)$ (blue), with $n=2$ and $\theta=\frac\pi3$, and the unit circle (red). Suitable values for the parameters $\alpha$ and $\beta$ exist if the curves intersect.}
  \label{curve}
\end{figure}

Such a $\beta$ exists if the curve of $f(\beta) = 1 - \left(1+e^{i\beta}\tan\frac\theta2\right)^n$ in the complex plane intersects the unit circle, as in Figure~\ref{curve}. Since $f$ is continuous, it suffices to exhibit one point outside the unit circle and one point within it. Consider
\begin{equation}
  f(0) = 1 - \left( 1+\tan\frac\theta2 \right)^n.
\end{equation}
Since $\tan\frac\theta2 \geq 2^{\frac1n} -1$, $f(0) \leq -1$, hence it is outside (or on) the unit circle. On the other hand,
\begin{equation}
  f(\pi) = 1 - \left( 1-\tan\frac\theta2 \right)^n.
\end{equation}
Since $0 \leq \tan\frac\theta2 \leq 1$, $0 \leq f(\pi) \leq 1$, hence it is inside (or on) the unit circle. This concludes the proof.

If the actual value of $\beta$ for a particular $\theta$ and $n$ is required, it is not difficult to find it numerically. For $n=2$, \eqref{betaeq} can even be solved analytically to find $\beta = \arccos\left(  (1 - 4t^2  - t^4) / 4t^3 \right)$ where $t = \tan\frac\theta2$.
\section{Formal, noise-tolerant version of the argument}\label{noiseversion}

This section proves Eq.~\eqref{finalresult}. This is a lower bound on the total variation distance between probability distributions corresponding to distinct quantum states, which holds even in the presence of noise. In the specific case of no noise ($\epsilon$ = 0), this section provides a more mathematical version of the argument already given in the main text.

Consider two methods of preparing a quantum system, such that quantum theory assigns the pure state $\ket{\psi_0}$ or $\ket{\psi_1}$. We assume that the quantum system after preparation has a real state $\lambda$. Each preparation method is associated with a probability distribution $\mu_i(\lambda)$ ($i=0,1$). This is to be thought of as the probability density for the system to be in the real state $\lambda$ after preparation. Another assumption is that when a measurement is performed, the behaviour of the measurement device depends only on the physical properties of the system and measuring device at the time of measurement. Formally, for a given measurement procedure $M$, the probability of outcome $k$ is given by $P(k| M, \lambda) = \xi_{M,k}(\lambda)$, where $\xi_{M,k}$ is a function $\xi_{M,k}: \Lambda\rightarrow [0,1]$. A model of this form reproduces the predictions of quantum theory exactly if
\begin{equation}
\int_\Lambda \xi_{M,k}(\lambda)\mu_i(\lambda)d\lambda = \bra{\psi_i} E_{M,k} \ket{\psi_i}, \label{correctprob}
\end{equation}
where $E_{M,k}$ is the positive operator which quantum theory assigns to outcome $k$.

The total variation distance between the distributions $\mu_0(\lambda)$ and $\mu_1(\lambda)$ is
\[
D(\mu_0,\mu_1) = \frac12\int_\Lambda \abs{\mu_0(\lambda) - \mu_1(\lambda)}d\lambda.
\]
The aim is to show that if a model of the above form reproduces the predictions of quantum theory approximately, so that for any measurement outcome, Eq.~\eqref{correctprob} holds to within $\epsilon$, then
\begin{equation}\label{mainresult}
  D(\mu_0,\mu_1) \geq 1 - 2\sqrt[n]{\epsilon}.
\end{equation} 
Eq.~\eqref{mainresult} holds for preparations of any pair of pure states $\ket{\psi_0}$ and $\ket{\psi_1}$, as long as $n$ is chosen to satisfy Eq.~\eqref{ppreq}.

To this end, consider $n$ independent preparations of quantum systems, where each can be chosen such that the quantum state is either $\ket{\psi_0}$ or $\ket{\psi_1}$. The joint quantum state is a direct product given by Eq.~(\ref{jointstate}). These systems will be brought together so that the joint measurement illustrated in Figure~\ref{fig2} and described in Section~\ref{eqnsol} can be performed. 

We have assumed that the behaviour of the measurement device is determined by its own properties, and by a complete list $\vec{\lambda} = (\lambda_1,\ldots,\lambda_n)$ of the real states of each one of the $n$ systems. Seeing as the systems are prepared independently, the probability distribution for $\vec{\lambda}$ is given by
\begin{equation}\label{productass}
\mu_{\vec{x}}(\vec{\lambda}) = \mu_{x_1}(\lambda_1)\times\cdots\times \mu_{x_n}(\lambda_n).
\end{equation}

In order to prove Eq.~(\ref{mainresult}), it is useful to define a quantity which we call the \emph{overlap} between $\mu_0(\lambda)$ and $\mu_1(\lambda)$:
\begin{equation}
\omega(\mu_0,\mu_1) = \int_\Lambda \min\{\mu_0(\lambda), \mu_1(\lambda)\}  d\lambda.\label{overlapdef}
\end{equation}
Note that $\omega(\mu_0,\mu_1) = 1 - D(\mu_0,\mu_1)$. 
For probability distributions $\mu_1, \dotsc, \mu_k$, the overlap can be generalised:
\begin{equation}
\omega(\mu_1,\ldots,\mu_k) = \int_\Lambda \min_i \mu_i(\lambda) d\lambda.
\end{equation}
Let $\Lambda^n$ denote the $n$-fold Cartesian product of $\Lambda$, i.e. $\Lambda^n$ is the space of possible values for $\vec{\lambda}$. From Eq.~(\ref{productass}), 
\begin{multline}
\min_{\vec{x}} \mu_{\vec{x}}(\lambda_1, \dotsc, \lambda_n) = \min\{\mu_0(\lambda_1),\mu_1(\lambda_1)\}\\\times \dotsm \times \min\{\mu_0(\lambda_n),\mu_1(\lambda_n)\}.
\end{multline}
Integrating both sides gives
\begin{equation}\label{multilem}
\omega\left( \{ \mu_{\vec{x}} \} \right) = \int_{\Lambda^n} \min_{\vec{x}} \ \mu_{\vec{x}}(\vec{\lambda}) \ \mathrm{d}\vec{\lambda} = \left(\omega(\mu_0,\mu_1)\right)^n.
\end{equation}


Now if the initial state is $\ket{\Psi(\vec{x})}$, and the measurement of Figure~\ref{fig2} of the main text is performed, Section~\ref{eqnsol} shows that the outcome corresponding to the basis state $\ket{\vec{x}}$ has probability zero according to quantum theory. If a model of the above form assigns probability $\leq \epsilon$ to this outcome, for any $\vec{x}$, then
\begin{equation}
\int_{\Lambda^n} \xi_{M,\vec{x}}(\vec{\lambda}) \mu_{\vec{x}}(\vec{\lambda}) \mathrm{d}\vec{\lambda} \leq \epsilon.
\end{equation}
Since $\min_{\vec{x}} \mu_{\vec{x}}(\vec{\lambda}) \leq \mu_{\vec{x}}(\vec{\lambda})$, and both $\xi_{M,\vec{x}}(\vec{\lambda})$ and $\mu_{\vec{x}}(\vec{\lambda})$ are non-negative,
\begin{equation}
\int_{\Lambda^n} \xi_{M,\vec{x}}(\vec{\lambda}) \min_{\vec{x}} \mu_{\vec{x}}(\vec{\lambda}) \mathrm{d}\vec{\lambda} \leq \epsilon.
\end{equation}
Finally, sum over $\vec{x}$ and use the normalization $\sum_{\vec{x}} \xi_{M,\vec{x}}(\vec{\lambda}) = 1$ to obtain
\begin{equation}\label{epsilonbound}
\omega\left( \{ \mu_{\vec{x}} \} \right) \leq 2^n \epsilon.
\end{equation}


Combining Eqs.~(\ref{multilem}) and (\ref{epsilonbound}) gives
\begin{equation}
(\omega(\mu_0,\mu_1))^n \leq 2^n \epsilon,
\end{equation}
which gives Eq.~(\ref{mainresult}).

\section{Numerical results}

\begin{figure}
  \begin{center}
\includegraphics[scale=0.7]{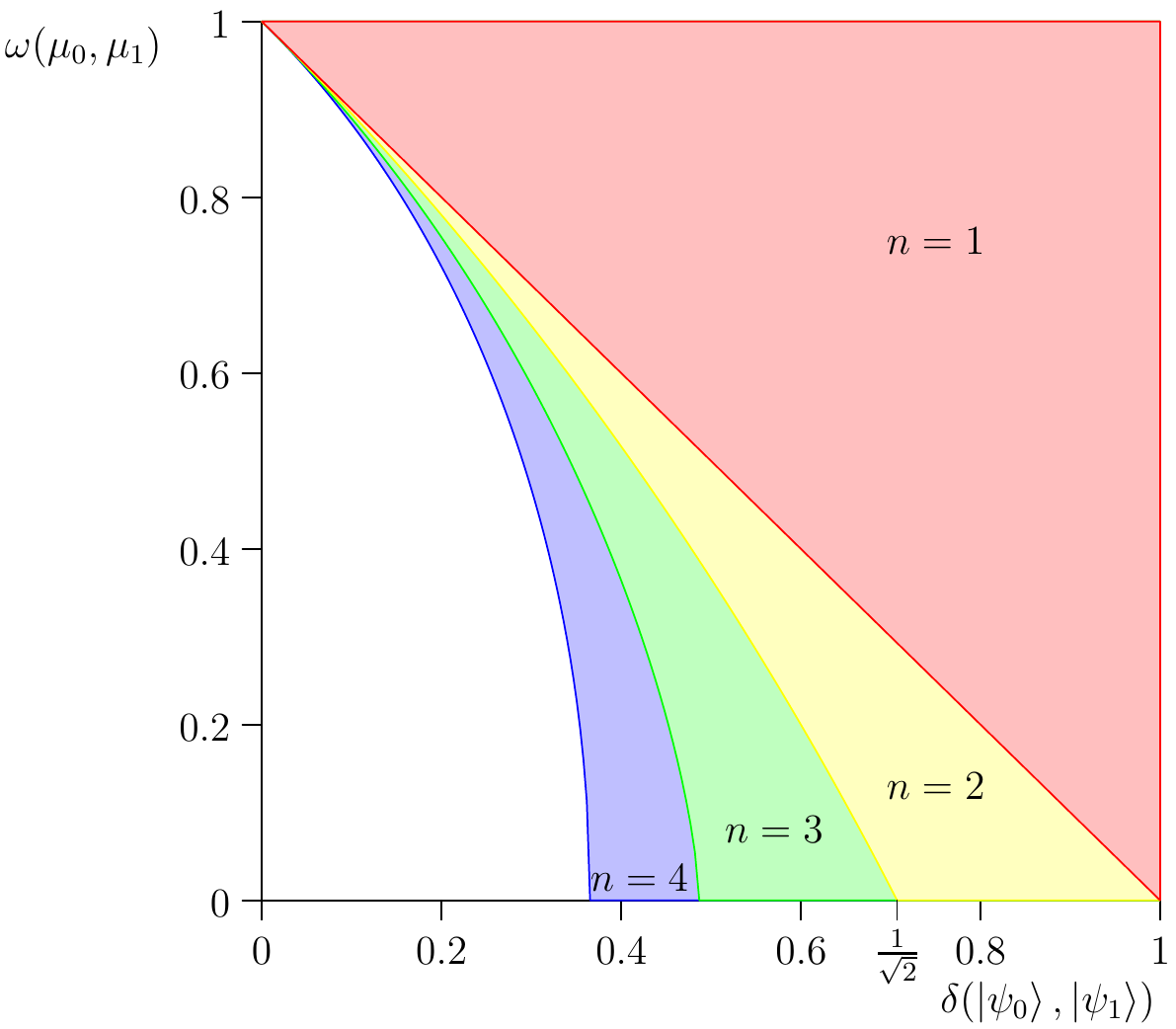}
  \end{center}
\caption{The overlap $\omega(\mu_0,\mu_1)$ (Equation~\eqref{overlapdef}), versus the quantum trace distance $\delta(\ket{\psi_0}, \ket{\psi_1}) = \sqrt{1 - \abs{\Braket{\psi_1|\psi_0}}^2}$. The red region is ruled out by measurements on a single system. The other regions can be ruled out by measurements on 2, 3 and 4 systems. The content of the no-go theorem is that larger and larger $n$ eventually fill the square, forcing $\omega(\mu_0, \mu_1) = 0$ for any pair of states. The boundaries of the regions are not ruled out (except that $\omega(\mu_0, \mu_1)>0$ \emph{is} ruled out for $\delta(\ket{\psi_0}, \ket{\psi_1}) = 1$).}
  \label{regions}
\end{figure}

For a given $\ket{\psi_0}$ and $\ket{\psi_1}$, the measurement described in Section \ref{eqnsol} requires the use of $n$ systems, with $n$ such that
\begin{equation}
  2\arctan\left(2^{1/n} - 1\right) \leq \arccos\abs{\Braket{\psi_0|\psi_1}}.\label{ppreq2}
\end{equation}
It is natural to ask if there exists a measurement that can make do with smaller values of $n$. We have checked for such a measurement by numerically solving \cite{sdpt,yalmip} the semi-definite program
\begin{equation}
\begin{aligned}
& \underset{E_i}{\text{minimize}}
& &\sigma := \sum_{\vec x} \tr(E_{\vec x}\Ket{\Psi(\vec x)}\Bra{\Psi(\vec x)}) \\
& \text{subject to}
&& E_{\vec x} \succeq 0,\\
&&& \sum_{\vec x} E_{\vec x} = \mathbb{I}.
\end{aligned}
\label{sdp}
\end{equation}
Since all the terms in the definition of $\sigma$ are non-negative, a measurement described by the POVM operators $\{E_{\vec x}\}$ can be used to prove the no-go theorem if and only if $\sigma = 0$. A variety of values of $\theta$ and $n$ were tested, and the minimum value of $\sigma$ was found to be $0$ exactly when $\eqref{ppreq2}$ is satisfied. Hence it appears that the measurement in Section \ref{eqnsol} uses the smallest possible number of systems.

Furthermore, when \eqref{ppreq2} is not satisfied the optimal measurement is of the form described in Section \ref{eqnsol}, but with $\alpha=\pi$ and $\beta=0$. (For $n=1$ this measurement is simply the standard minimum error discrimination measurement for $\ket{\psi_0}$ and $\ket{\psi_1}$.) By a similar argument to the previous section, if the quantum theory predictions for this measurement hold, then $(\omega(\mu_0,\mu_1))^n \leq \sigma$. Hence, in addition to our main result that there exists a measurement showing $\omega(\mu_0, \mu_1) = 0$ when \eqref{ppreq2} is satisfied, this measurement can be used to place bounds on $\omega(\mu_0, \mu_1)$ when it is not. The situation is depicted in Figure~\ref{regions}. 

Finally, we note that the problem \eqref{sdp} has an unusual operational interpretation. By considering each outcome $E_{\vec x}$ as the identification of $\Ket{\Psi(\vec x)}$, we have an error probability of $1 - \sigma/2^n$, and so this is the ``maximum error'' discrimination problem for the quantum states $\{\Ket{\Psi(\vec x)}\}$ (with equal priors). For the special cases of two states this becomes the minimum error problem under swapping of the outcome labels.

\end{document}